# Dual-gated graphene devices for near-field nano-imaging

Sai S. Sunku[1,2,*], Dorri Halbertal[1,*], Rebecca Engelke[4], Hyobin Yoo[4], Nathan R. Finney[3], Nicola Curreli[3], Guangxin Ni[1], Cheng Tan[3], Alexander S. McLeod[1], Chiu Fan Bowen Lo[1], Cory R. Dean[1], James C. Hone[3], Philip Kim[4], D. N. Basov[1],†

[1] Department of Physics [2] Department of Applied Physics and Applied Mathematics [3] Department of Mechanical Engineering, Columbia University, New York, NY

[4] Department of Physics, Harvard University, Cambridge, MA

* These authors contributed equally

† db3056@columbia.edu

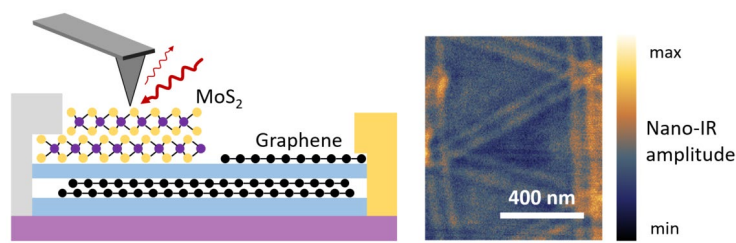

## Abstract

**Graphene-based heterostructures display a variety of phenomena that are strongly tunable by electrostatic local gates. Monolayer graphene (MLG) exhibits tunable surface plasmon polaritons, as revealed by scanning nano-infrared experiments. In bilayer graphene (BLG), an electronic gap is induced by a perpendicular displacement field. Gapped BLG is predicted to display unusual effects such as plasmon amplification and domain wall plasmons with significantly larger lifetime than MLG. Furthermore, a variety of correlated electronic phases highly sensitive to displacement fields have been observed in twisted graphene structures. However, applying perpendicular displacement fields in nano-infrared experiments has only recently become possible (*1*). In this work, we fully characterize two approaches to realizing nano-optics compatible top-gates: bilayer $MoS_2$ and MLG. We perform nano-infrared imaging on both types of structures and evaluate their strengths and weaknesses. Our work paves the way for comprehensive near-field experiments of correlated phenomena and plasmonic effects in graphene-based heterostructures.**

Graphene-based van der Waals (vdW) heterostructures display a variety of phenomena including superior plasmonic properties (*2–5*), tunable band structures (*6–8*), topological edge states (*9–11*), and correlated phases such as superconductivity (*12*, *13*). This large variety of electronic phases arises because the properties of graphene are strongly tunable by electrostatic gates. The optical excitations corresponding to these phases lie in the infrared range of the electromagnetic spectrum (*14*), where the wavelength of light, $\lambda_0$, ranges from 1µm to 100µm. Probing such heterostructures with conventional far-field optical experiments is challenging because of their limited lateral dimensions compared to $\lambda_0$. However, tip-based scanning nano-infrared experiments can overcome the diffraction limit and achieve a spatial resolution better than 10nm (*15*).

Nano-infrared experiments have established monolayer graphene (MLG) as an excellent platform for plasmonics because of a large confinement ratio $\lambda_0/\lambda_p$ ($\lambda_p$ is the plasmon wavelength) (*16–18*), tunability with an external gate (*3*, *19*) and long lifetimes for the SPPs approaching 2 ps (*4*, *20*). While MLG is well studied, the plasmonic properties of bilayer graphene are relatively unexplored (*21*). When bilayer graphene is gapped and the Fermi level lies in the gap, exotic plasmonic phenomena are predicted to occur. Gapped BLG under photoexcitation is predicted to amplify SPPs (*22*) while domain wall solitons in gapped BLG could host one-dimensional SPPs with lifetimes approaching $10^2$ ps (*23*).

Nano-infrared experiments have also begun to probe multilayer graphene-based Moiré systems that are known to host correlated electronic phenomena such as twisted bilayer graphene (TBG) (*12*), twisted trilayer graphene (TTG) (*24*) and twisted double bilayer graphene (TDBG) (*25*, *26*). The electronic properties of all these systems are strongly sensitive to perpendicular displacement field. For example, the correlated insulator phases in TTG and TDBG appear only for a limited range of displacement fields (*24*, *25*).

In transport experiments, perpendicular displacement field can be introduced using an evaporated metal layer or a graphite layer as a top-gate in conjunction with a back-gate. While transport (*27*) and some far-field optical experiments (*28–30*) can be performed on such structures, they are incompatible with nano-infrared experiments for multiple reasons. First, such layers are relatively thick (tens of nm) which make the underlying graphene layer inaccessible to nano-optics experiments. When the layers are made thinner, the presence of a high density of high mobility free carriers leads to plasmonic effects in the top-gate which modifies and obscures the behavior of the underlying graphene layer.

Recent work has shown that MLG could be used as a top gate to study Moiré patterns in vdW heterostructures (*1*). However, the capabilities and limitations of the top gate were not fully explored. While Ref (*1*) showed that two-dimensional domains in a Moiré pattern could be visualized through a MLG top gate, it's not yet known if the plasmonic phenomena in the underlying graphene layer and one-dimensional features such as domain walls in BLG can be resolved. In this work, we demonstrate and fully characterize two approaches for nano-optics compatible top gates: bilayer MoS₂ and MLG. We are able to visualize the plasmons in the MLG and TBG layers underneath the MoS₂ top-gate. We further demonstrate a depletion of the carrier density of the underlying graphene layers with the top-gate through measurements of the plasmon wavelength and nano-infrared scattering amplitude. We then explore the use of a MLG layer as the top-gate for BLG. The doped MLG layer is a robust top-gate, but has strong optical response at mid-infrared frequencies of its own. We therefore explore the possibility of selectively probing the underlying BLG through nano-photocurrent imaging and are able to visualize domain walls in the BLG layer. Our work paves the way for realization of fully tunable vdW devices compatible with nano-optics experiments.

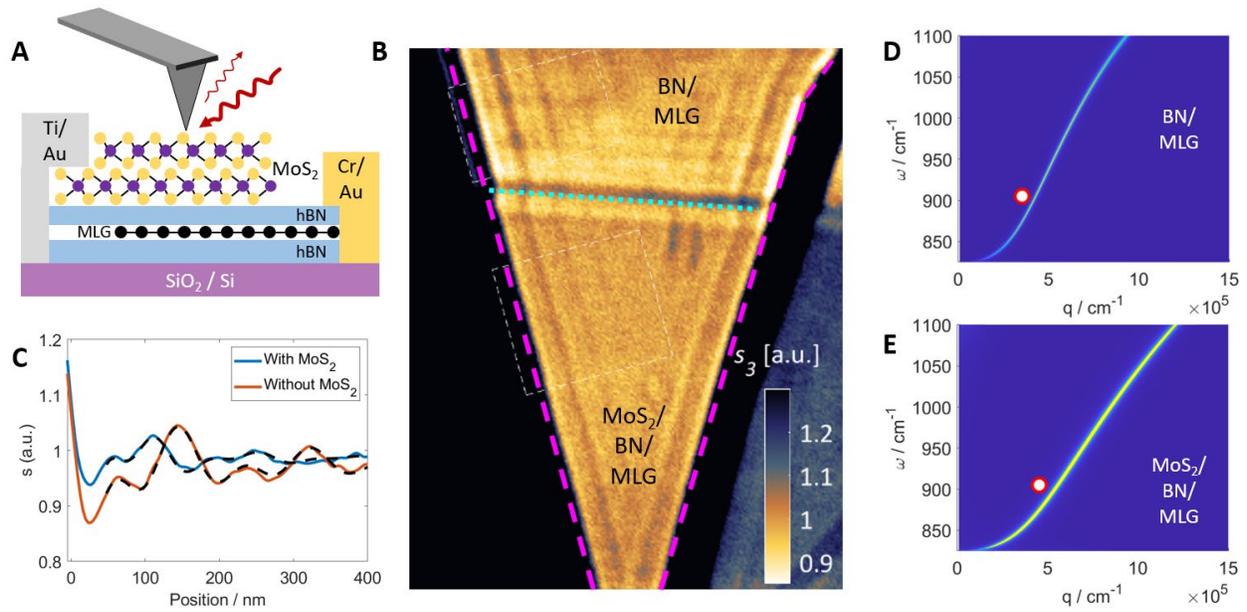

**Figure 1 | Nano-optics measurements on MoS₂-gated monolayer graphene.** (A) Schematic of our nano-infrared experiment and our device. Bilayer MoS₂ is contacted by Ti/Au from above while the graphene has a side contact made of Cr/Au. (B) Two-dimensional image of the nano-infrared amplitude $s$ over our device for $V_{bg} = +80V$ and $V_{tg} = 0V$. The edges of MLG and MoS₂ are represented by magenta dashed and green dotted line represents respectively. (C) Line profiles of the nano-infrared amplitude $s$ across a graphene edge showing plasmon polaritons. Black dashed lines represent fits to a damped oscillations model (Section S3 of Supporting Information). (D) Calculated imaginary part of the reflection coefficient $\text{Im}(r_P)$ matching the two experimental heterostructures: (D) BN/MLG/BN/SiO₂ (E) MoS₂/BN/MLG/BN/SiO₂. The bright contour of maximal values corresponds to the plasmon mode. The circles in (D) and (E) correspond to experimental data extracted from panel (C) (Section S3 of Supporting Information).

Figure 1(A) shows a schematic of our experimental setup. Our first device consists of monolayer graphene encapsulated between hexagonal boron nitride (hBN) layers. The thickness of the top hBN layer is kept small (2 nm) to allow optical near-field access to the underlying graphene layer. A bilayer of $MoS_2$ is then placed on the top hBN layer for use as a top-gate while a doped silicon layer underneath the heterostructure serves as the bottom-gate. We chose $MoS_2$ because it is expected to be transparent to mid-infrared light. We study this device using a scanning nano-infrared microscope where incident light from a quantum cascade laser is focused onto the apex of a sharp metallic tip. We used light of frequency $\omega = 1/\lambda_0 = 905$ cm$^{-1}$ for all experiments in this manuscript. The amplitude $s$ and phase $\phi$ of the scattered light are detected with an interferometric method (*31*). The sharpness of the tip provides the momentum necessary to launch SPPs which propagate radially outward from the tip. When the SPPs encounter a physical (*3*, *19*) or electronic (*32*) boundary, they are reflected and form a standing wave pattern that we directly visualize.

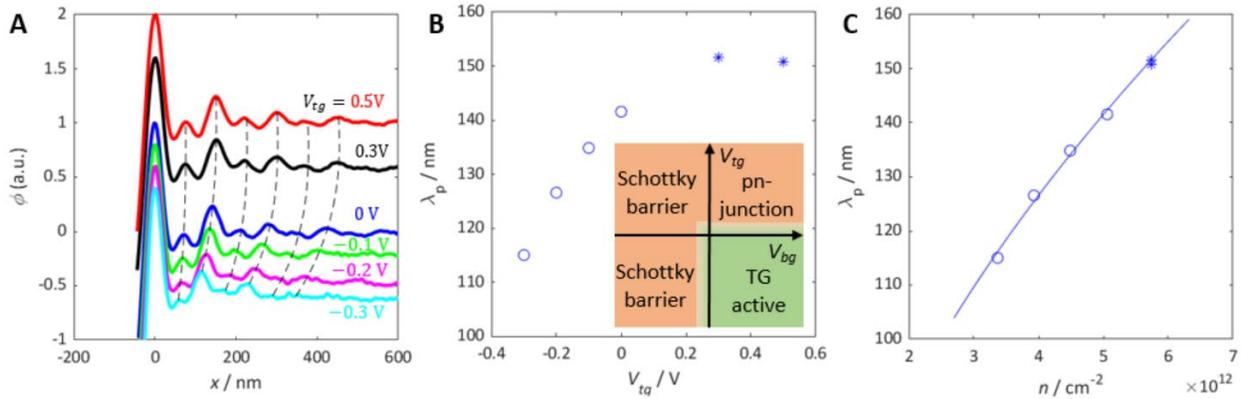

**Figure 2 | Carrier density modulation in monolayer graphene with MoS$_2$ top-gate.** (A) Line profiles of the nano-infrared phase $\phi$ for various values of $V_{tg}$ at $V_{bg} = +80$V, showing a clear change in the plasmon wavelength. The data is scaled and shifted for clarity. Dashed lines follow the plasmonic peaks and are used to extract $\lambda_p$. (B) Plasmon wavelength as a function of the top-gate voltage $V_{tg}$. Circles represent $V_{tg} \leq 0$V and asterisks represent $V_{tg} > 0$V. Inset shows the behavior of the MoS$_2$ top-gate for various $V_{tg}$ and $V_{bg}$ values. The top-gate is most effective in one of the four quadrants and its performance decays quickly in other quadrants. (C) $\lambda_p$ as a function of the estimated carrier density in the graphene layer $n$. The data points for $V_{tg} > 0$V cluster together because the top-gate is ineffective (described in text).

Figure 1(B) shows a two-dimensional map of the nano-infrared amplitude $s$ measured on our device with $V_{bg} = +80$V and $V_{tg} = 0$V. The green dotted line marks the edge of the MoS$_2$ layer such that the area above the line does not contain MoS$_2$. We observe clear fringes parallel to the edges of the MLG (marked by magenta dashed lines) throughout the image. These fringes confirm that we are able to launch and image SPPs in the MLG layer even when the MLG is underneath MoS$_2$.

A comparison of the fringes above and below the MoS$_2$ boundary in Fig 1(B) indicates that the plasmon wavelength is smaller in the region with MoS$_2$. In Fig 1(C), we plot the line profiles extracted across the graphene edge from both regions. The line profiles confirm that the plasmon wavelength is reduced to 138 nm under the MoS$_2$ layer (blue line in Fig 1(C)) compared to 177 nm without MoS$_2$ (orange line in Fig 1(C)). This reduction is due to the large static dielectric constant of MoS$_2$ and the resulting screening. This change in plasmon wavelength is consistent with the calculated change in plasmon dispersion (Fig 1(D) and 1(E)).

Figure 2 demonstrates the tuning of carrier density in the MLG layer with the MoS$_2$ top-gate. Figure 2(A) shows line-profiles of the nano-infrared phase $\phi$ for different values of top-gate bias $V_{tg}$ for a fixed value of back-gate bias $V_{bg} = +80\text{V}$. We observe a clear change in the plasmon wavelength as $V_{tg}$ is changed. At negative values of $V_{tg}$, we observe a decrease in the plasmon wavelength which is consistent with a depletion of the carrier density in the graphene layer. When $V_{tg}$ is tuned to +0.3V, we observe an increase in $\lambda_p$. But a further increase in $V_{tg}$ to +0.5V does not change $\lambda_p$, indicating that the carrier density in MLG does not change (Figure 2(B) and 2(C)). This limitation is the result of a pn-junction forming in the MoS$_2$ layer as described below. Taken together, our results confirm that we are able to deplete the carrier density in the graphene layer which is necessary for realizing gapped BLG.

We now turn to the BLG region of our heterostructure that is also covered by the MoS$_2$ top-gate. The BLG in our heterostructure was produced by a 'tear-and-stack' technique (Section S1 of Supporting Information) which resulted in a small twist angle (estimated to be ~0.02°) between the layers and a large Moiré pattern. Atomic relaxation leads to the formation of larger domains of Bernal bilayer graphene separated by domain walls (*33*) that host topological states (*9–11*). The change in optical conductivity arising from the topological states reflects plasmon polaritons leading to fringes in nano-infrared experiments (*34, 35*). Changing the carrier density and interlayer bias across the BLG changes the optical conductivity across the domain wall and modifies the fringe pattern (*34*). Figure 3(B) shows the nano-infrared amplitude over a region containing several domain walls for $V_{bg} = +80\text{V}$ and $V_{tg} = 0\text{V}$. We observe features in the amplitude that correspond to plasmons reflecting off the domain walls (*35, 36*). As we increase $V_{tg}$, we observe a clear change in the plasmonic pattern that confirms the changing carrier density and interlayer bias in the BLG layer. By demonstrating dual-gating and observing propagation of plasmons, we have thus shown the feasibility of performing nano-infrared studies of a dual gated system using this approach.

Next, we discuss the limitations of the TMD top-gate. First, we consider the performance of the top-gate at a negative $V_{bg}$. Because of the high work function of MoS$_2$, evaporated metals typically make n-type contact to MoS$_2$ (*37*). The geometry of our device is such that the titanium metal contacts to the MoS$_2$ lie outside the graphene region (Fig 1A). Therefore the contact resistance at the Ti/MoS₂ layer depends only on $V_{bg}$. At a large negative $V_{bg}$, the Schottky barrier at the Ti/MoS₂ junction is too large and renders the top-gate ineffective. Therefore, unless doped by local gates (*38*), the MoS₂ top-gate is only functional for positive $V_{bg}$ where n-type carriers are injected into the MoS₂ layer.

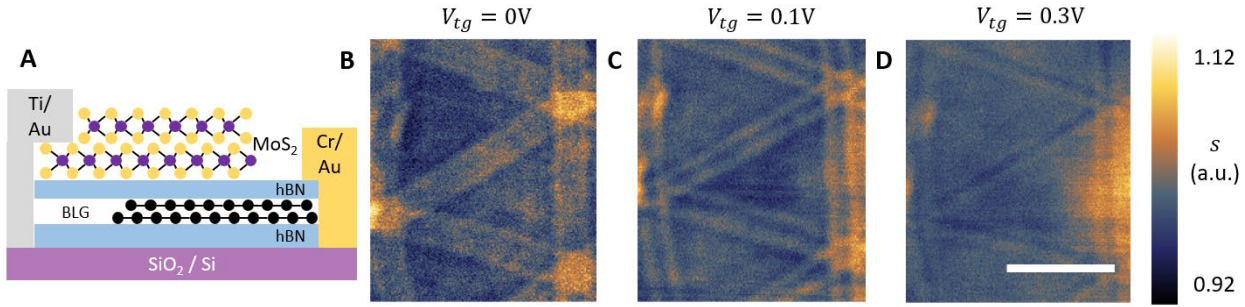

**Figure 3 | Demonstration of top-gating effect in bilayer graphene domain walls.** (A) Schematic of the heterostructure. (B, C, D) Nano-infrared amplitude image of domain walls in BLG for three different top-gate voltages with $V_{BG} = +80\text{V}$. Scale bar 400nm.

At a fixed, positive $V_{bg}$, the region by the contacts remains n-doped and the application of $V_{tg}$ starts to change the carrier density in the MoS₂ region directly above the graphene layer. When $V_{tg}$ is negative, the carriers in the MoS₂ layer are all n-type. However, as $V_{tg}$ becomes positive, the MoS₂ region above the graphene becomes hole-like. Since the carriers close to the contacts remain electron-like, a pn-junction forms in the MoS₂ layer along the graphene edge. This pn-junction isolates the Ti contacts from the MoS₂ region above the graphene layer and causes the top-gate to become ineffective. Taking the effects of the Schottky barrier and the pn-junction together, we conclude that the top-gate is most effective only in one of the four quadrants in the $V_{tg} - V_{bg}$ plane and its performance decays quickly in the other quadrants, as illustrated in the inset of Fig 2(B).

To achieve full control over the properties of BLG, we could consider other materials as a top-gate. While a p-type TMD such as WS₂ can lead to a top-gate that is functional at negative $V_{bg}$, the pn-junction limitation would still restrict its functionality to just one quadrant in the $V_{bg} - V_{tg}$ plane. This limitation arises directly because of the electronic bandgap and therefore would be present for any semiconductor. Only a gapless ambipolar material, such as monolayer graphene, can overcome this limitation.

We now explore the possibility of using monolayer graphene (MLG) as a top-gate for BLG. The ambipolar nature of MLG means that the contacts do not restrict the range of operational gate voltages. However, MLG has a strong optical response of its own in mid-infrared frequencies. At the same time, if the Fermi energy in the top gate layer is very small, interband transitions in the top gate layer will lead to an increased damping that can obscure the plasmonic features in nano-infrared imaging (Section S4.1 of Supporting Information). Therefore, we also explored the nano-photocurrent technique which can selectively probe the underlying BLG layer.

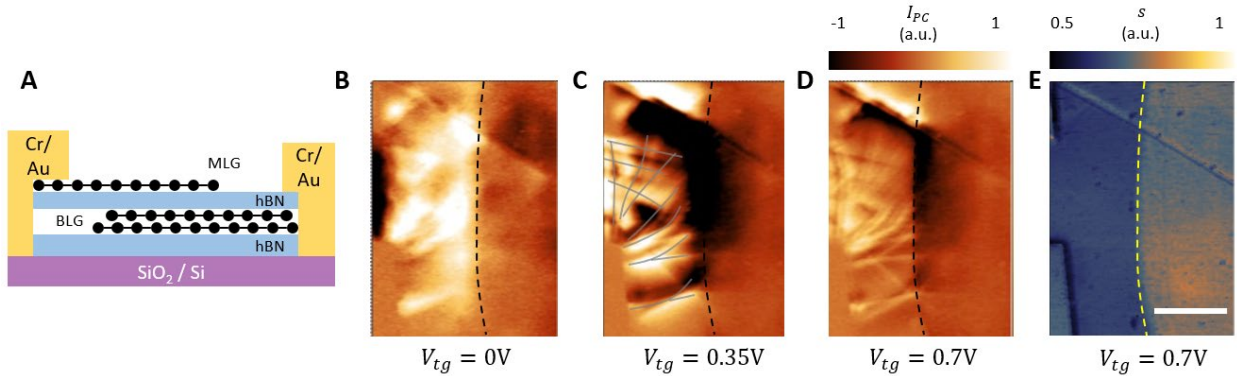

**Figure 4 | Monolayer graphene as a top-gate for domain walls in bilayer graphene.** (A) Schematic of the heterostructure (B, C, D) Nano-photocurrent images for three different $V_{tg}$ showing domain walls in the underlying bilayer graphene. (E) Nano-infrared amplitude image at $V_{tg} = 0.7\text{V}$. Black dashed lines in (B-D) and yellow dashed line in (E) correspond to the boundary of the top-gated region. Grey solid lines in (C) indicate the domain walls in BLG. Scale bar 1μm.

Figure 4(A) shows a schematic of our heterostructure with a MLG top gate. Figure 4(E) shows the nano-infrared amplitude image of our device with $V_{bg} = 0\text{V}$ and $V_{tg} = +0.7\text{V}$. The yellow dashed line indicates the boundary of the MLG top gate. We observe a nano-infrared contrast indicating that the top gate is active but we see no other features, most likely because the carrier density in the bilayer graphene is too low (Section S1.4 of Supporting Information). Figure 4 (A-D) shows the results of nano-photocurrent experiments (*39*) at different $V_{tg}$. As $V_{tg}$ is increased from zero, the nano-photocurrent begins to resemble the photocurrent profiles seen in other twisted BLG heterostructures (*40*, *41*) and are known to arise from domain walls. Based on the periodicity of the Moiré pattern, we estimate a twist angle of ~0.1°. The irregularity of the domain wall pattern in Fig 4 in comparison to Fig 3 is due to strain accumulated during the fabrication process. These results demonstrate that we are able to resolve the domain wall pattern in nano-photocurrent through a doped MLG top gate.

Finally, we directly compare the properties of MLG and bilayer MoS$_2$ top gates for nano-infrared experiments with the following hypothetical scenario. We consider an encapsulated heterostructure of monolayer graphene with a carrier density of $n = 6 \cdot 10^{12} \text{cm}^{-2}$ with either a MLG top gate (Fig. 5A) or a TMD top gate (Fig. 5B). We then vary the carrier concentration only in the top gate to observe how strongly the top gate modifies the behavior of the underlying graphene layer. The plasmonic dispersions in both cases are shown in Figure 5. The dispersion changes significantly with a MLG top gate while it remains mostly unchanged in case of the TMD top gate. The large change in the dispersion with the MLG top gate is due to strong hybridization between the plasmonic modes in the two graphene layers (*42–44*). The smaller change in dispersion with a TMD top gate demonstrates that the hybridization of the plasmonic modes is negligible with a TMD top gate. The thickness of the TMD top gate leads to a small but significant effect on the plasmonic dispersion, as discussed in Section S5 of Supporting Information. These results suggest that a TMD top gate allows direct access to the plasmonic phenomena in the underlying graphene layer in nano-infrared experiments with minimal obscuring.

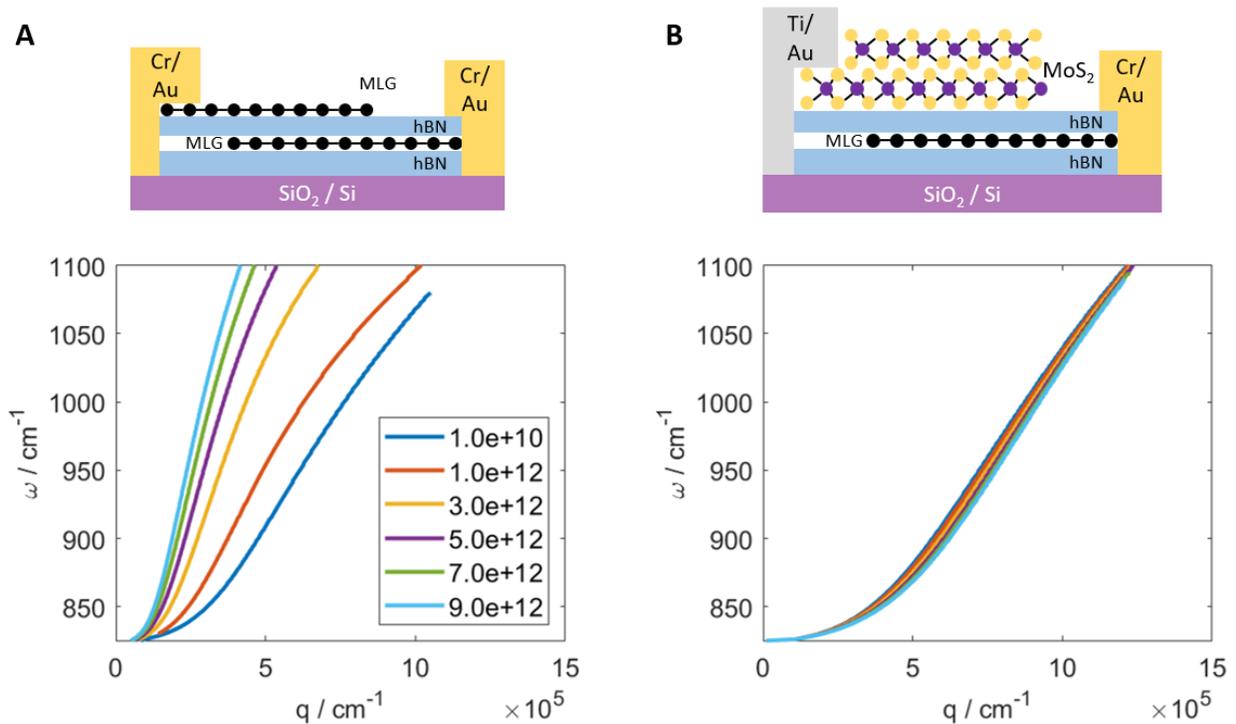

**Figure 5 | Direct comparison between the MLG and TMD top gates.** (A) Change in the dispersion of the plasmonic mode of the heterostructure as the carrier density in the top gate is varied for a MLG top gate. (B) Similar plot as (A) for bilayer MoS$_2$ top gate. The thicknesses of the top and bottom hBN are 5nm and 30nm respectively and the underlying MLG layer is doped to $n = 6 \cdot 10^{12} \text{cm}^{-2}$.

In conclusion, our results demonstrate two near-field compatible top gates for BLG. With MoS$_2$, we were able to study plasmons through scattering nano-infrared experiments and demonstrate the depletion of carriers in the underlying graphene layers. With a MLG top gate, we were able to visualize the domain walls in the underlying BLG through nano-photocurrent experiments. Our work paves the way for exploring the plasmonic properties of gapped bilayer graphene with scanning nano-infrared and nano-photocurrent experiments.

**Acknowledgements**: Research in van der Waals heterostructures at Columbia was solely supported as part of Programmable Quantum Materials, an Energy Frontier Research Center funded by the U.S. Department of Energy (DOE), Office of Science, Basic Energy Sciences (BES), under award DE-SC0019443. D.N.B. is the Vannevar Bush Faculty Fellow (N00014-19-1-2630) and Moore Investigator in Quantum Materials EPIQS #9455. D.H. was supported by a grant from the Simons Foundation (579913). The work at Harvard was supported by NSF DMREF (DMR-1922172). N.R.F. acknowledges support from the Stewardship Science Graduate Fellowship program provided under cooperative agreement number DE-NA0003864. N.C. acknowledges the project SONAR, which has received funding from the European Union's Horizon 2020 research and innovation programme under the Marie Skłodowska–Curie grant agreement (no. 734690).

# Supporting Information for "Dual-gated graphene devices for near-field nano-imaging"

Sai S. Sunku[1,2,*], Dorri Halbertal[1,*], Rebecca Engelke[4], Hyobin Yoo[4], Nathan R. Finney[3], Nicola Curreli[3], Guangxin Ni[1], Cheng Tan[3], Alexander S. McLeod[1], Chiu Fan Bowen Lo[1], Cory R. Dean[1], James C. Hone[3], Philip Kim[4], D. N. Basov[1],[†]

[1] Department of Physics [2] Department of Applied Physics and Applied Mathematics [3] Department of Mechanical Engineering, Columbia University, New York, NY

[4] Department of Physics, Harvard University, Cambridge, MA

* These authors contributed equally

† db3056@columbia.edu

## Section S1: Methods

### S1.1 Device fabrication

The results shown in Figs 1-3 were obtained from a $MoS_2$ top-gated device (Device 1) consisting of graphene layers encapsulated in hexagonal boron nitride with a $MoS_2$ layer on top. The stack was fabricated using the dry transfer method. A poly(bisphenol A carbonate) (PC) coated on a stamp made of transparent elastomer polydimethylsiloxane (PDMS). The two graphene layers to form the bilayer graphene were assembled by tearing a large single layer of graphene and stacking them together. The inherent strain in this process results in one of the layers twisting slightly relative to the other layer and leads to the formation of domain walls (S*1*). The contacts to the TMD layer were made of titanium while side contacts (S*2*) to the graphene layer were made of chromium and gold.

The MLG top-gated device (Device 2) was also fabricated with the dry transfer method but with a poly(propylene carbonate) (PPC) coated stamp. The heterostructure was fabricated in the reverse order so that the MLG top-gate was not contaminated by contact with the PPC polymer and flipped onto a $SiO_2$/Si chip. After flipping, the heterostructure was annealed in vacuum to remove the PPC residue. Electrical contacts to the graphene layers were made with chromium and gold.

The results shown in Fig S9 were obtained from another $MoS_2$ top-gated device (Device 3) that was also fabricated with the dry transfer method.

## S1.2 Infrared nano-imaging

Infrared nano-imaging was performed with a commercial scattering-type scanning near-field optical microscope (s-SNOM) based on a tapping mode atomic force microscope from Neaspec GmbH. Our light source was a quantum cascade laser obtained from DRS Daylight Solutions, tunable from 900 cm$^{-1}$ to 1200 cm$^{-1}$. The light from the laser was focused onto a metallic tip oscillating at a tapping frequency of around 250 kHz with a tapping amplitude of around 60 nm. The scattered light was detected using a liquid nitrogen cooled HgCdTe (MCT) detector. To suppress far-field background signals, the detected signal was demodulated at a harmonic $n$ of the tapping frequency. In this work, we used $n = 3$.

## S1.3 Plasmon wavelength and dispersion calculations

The dielectric constants for MoS$_2$, hBN, and SiO$_2$ used in the reflection coefficient calculations of Fig 1(D), 1(E), and the rest of the manuscript were obtained from Refs (S*3*), (S*4*), and (S*5*) respectively. The plasmon wavelength used to plot the crosses was determined by the spacing between the fringes in the spatial profiles of the near-field phase (Fig 2(A)).

For the calculations in Figure 5, we assumed that the dielectric properties of a doped MoS$_2$ layer can be described by a Drude model,

$$\epsilon = \epsilon_\infty + \frac{\omega_p^2}{\omega^2 - i\gamma\omega}$$

$$\omega_p^2 = \frac{ne^2}{m^*\epsilon_\infty}$$

where $\epsilon_\infty$ and $\epsilon_0$ are the high-frequency and low-frequency dielectric constants, $\omega_p$ is the plasma frequency, $\gamma$ is the damping, $n$ is the carrier density and $m^*$ is the effective mass of the carriers and $e$ is the electron charge. We obtained values for $\epsilon_\infty$ from Ref (S*3*): along ab-plane $\epsilon_\infty = 15.2$ and along the c-axis $\epsilon_\infty = 6.2$. The band structure of TMDs is anisotropic with the out-of-plane effective mass expected to be smaller than the in-plane effective mass but is not known accurately. Here, we assumed that the effective mass $m^*$ was isotropic as a worst case scenario and equal to the measured in-plane effective mass $m^* = 0.45\ m_0$ (S*6*), , where $m_0$ is the free electron mass. The plasmon dispersion is insensitive to $\gamma$ and we used $\gamma = 300 \text{cm}^{-1}$, based on Ref (S*7*).

## S1.4 Bilayer graphene parameter estimates

In this section, we calculate the Fermi energy and interlayer bias of bilayer graphene for the heterostructures considered in this work. The top and bottom gates produce displacement fields above and below the graphene layer given by $D_t = \epsilon_t V_{tg}/d_t$ and $D_b = \epsilon_b V_{bg}/d_b$ where $\epsilon_t$, $\epsilon_b$ and $d_t$, $d_b$ are the dielectric constant and thickness of the top and bottom gate dielectrics respectively (Fig S1(a)). The band structure is affected by a combination of the two displacement fields, as shown schematically in Fig S1(b). The carrier density in the graphene layer, $n$, is determined by the difference $D_b - D_t$ and the band gap $\Delta$ is determined by $\bar{D} = (D_b + D_t)/2$ (S8, S9). We used Ref (S9) to convert $\bar{D}$ to $\Delta$ and computed the Fermi energy $E_F$ by varying it until the carrier density computed with a tight binding model (S10) matched the desired carrier density. The results are summarized in Table 1.

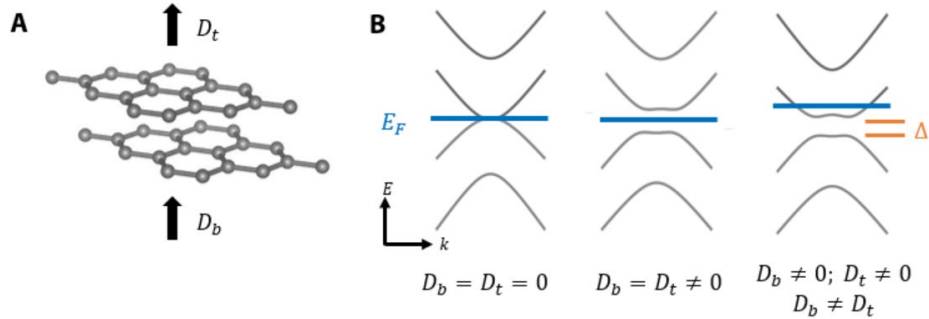

**Figure S1 | Band structure of bilayer graphene for different displacement fields.** (A) Schematic of bilayer graphene with top and bottom displacement fields, $D_t$ and $D_b$, arising from the top and bottom gates. (B) Schematic band structures for various combinations of $D_b$ and $D_t$. The bandgap, represented by $\Delta$, is determined by $\bar{D} = (D_b + D_t)/2$.

| | |
|---|---|
| $n$ as a function of $D_b - D_t$ | $n = \dfrac{\epsilon_0}{e}(D_b - D_t) * 1 \cdot 10^5$ |
| $V_i$ as a function of $\bar{D}$ | $\Delta = 105\,\bar{D} + 11.1\,\bar{D}^2 - 6.36\,\bar{D}^3$ (based on Ref (S9)) |
| $E_F$ as a function of $n$ | $E_F = 1.97 \cdot 10^{-14}\,n - 5.32 \cdot 10^{-28} n^2 + 1.00 \cdot 10^{-41}\,n^3$ (based on the tight binding model of Ref (S10)) |

**Table 1 | Polynomial fits for calculating the parameters of BLG for a set of gate voltages.** $D_b$, $D_t$ and $\bar{D}$ are in units of V/nm, $n$ is in units of cm$^{-2}$, and $\Delta$ and $E_F$ are in units of meV.

The gate voltages used in Figures 4(D) and 4(E) correspond to $D_t = 0.4$ V/nm and $D_b = 0$. Based on the fits in Table 1, we see that the corresponding BLG parameters are $\Delta = 21.4$ meV and $n = 2.21 \cdot 10^{12}$ cm$^{-2}$. These values correspond to a Fermi energy $E_F = 41$ meV. Given that our probing energy $\omega = 905$ cm$^{-1}$ is equivalent to 112 meV, the Fermi energy is too low to prevent interband transitions which occur at frequencies $\omega > 2E_F$. Therefore, we cannot exclude the possibility that the reason we see no nano-infrared contrast from the domain walls in Figure 4(E) is because of the increased damping in the BLG layer.

# Section S2: Nano-infrared images at different top-gate voltages

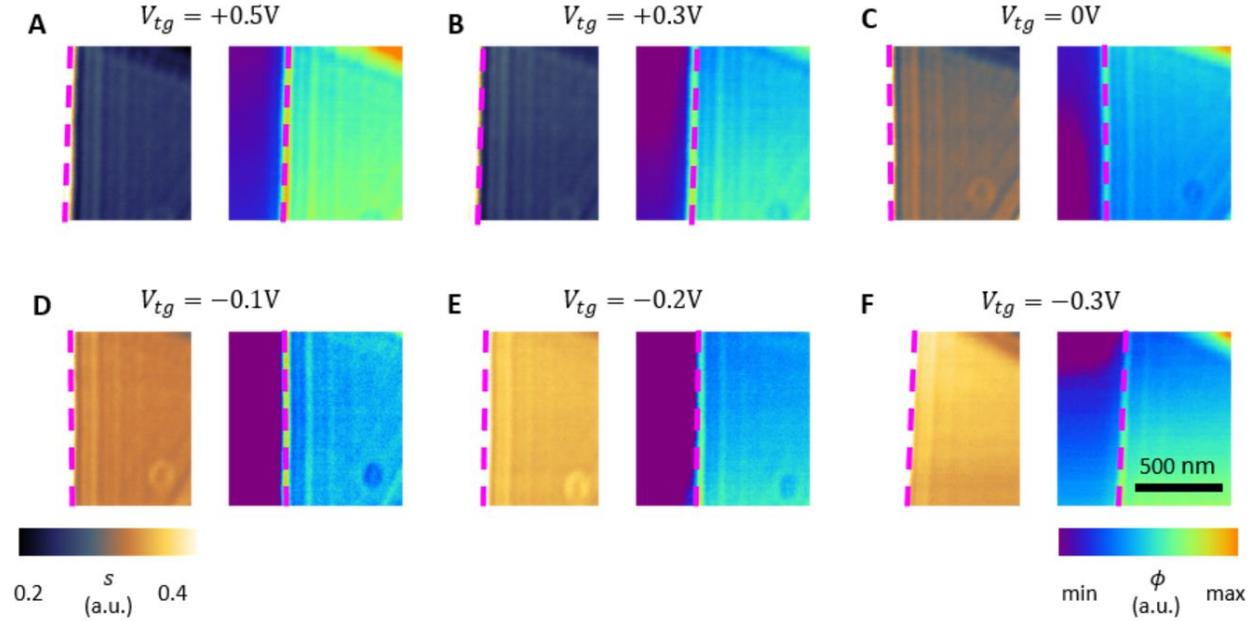

**Figure S2 | Nano-infrared images of monolayer graphene (MLG) at $V_{bg} = +80V$ and various values of $V_{tg}$.** The images on the left are the nano-infrared amplitude $s$ while the images on the right are of the nano-infrared phase $\phi$. The magenta lines represent the physical boundary of the MLG. The color bar limits for the amplitude images are identical for all gate voltages. The variation in the nano-infrared amplitude is consistent with a decrease (increase) of the carrier concentration as $V_{tg}$ is decreased (increased) (S*11*).

# Section S3: Line profile fits using the damped oscillations model

The nano-infrared amplitude line profiles were fit using the damped oscillations model (S*12*):

$$s = \frac{\cos(2qx)\exp(-2q\gamma x)}{\sqrt{x}} + \alpha\frac{\cos(qx+\phi)\exp(-q\gamma x + \phi)}{x}.$$

Here, the first term represents plasmons that are launched by the tip and reflected by the graphene edge and the second term represents the plasmons that launched by the edge. The edge is assumed to be located at located at $x = 0$. $q$ is the plasmon momentum defined to be $q = \frac{2\pi}{\lambda_p}$, $\gamma$ is the damping of the plasmonic wave, and $\alpha$ and $\phi$ capture the difference in the magnitude and phase between the tip-launched waves and the edge-launched waves. The obtained fits are shown in Figure S2(A) and Figure 1(C) of the main text. The parameters derived from the fits are shown in Figure S2(B).

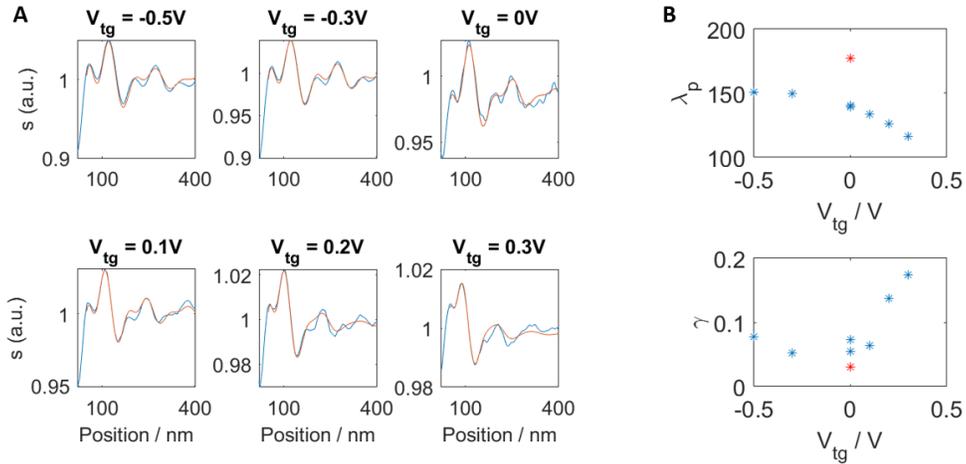

**Figure S3 | Damped oscillations model fits for different top-gate voltages.** (A) Nano-infrared amplitude line profiles for different top-gate voltages. (B) The plasmon wavelength $\lambda_p$ and the damping $\gamma$ extracted from the fits. The red asterisks correspond to the data without MoS$_2$ (red curve in Figure 1(C) of main text).

### Section S4: Nano-infrared vs nano-photocurrent with an MLG top gate

In this section, we explore in detail the advantages and disadvantages of nano-infrared imaging and nano-photocurrent techniques with a monolayer graphene (MLG) top gate.

### Section S4.1: Nano-infrared imaging simulations

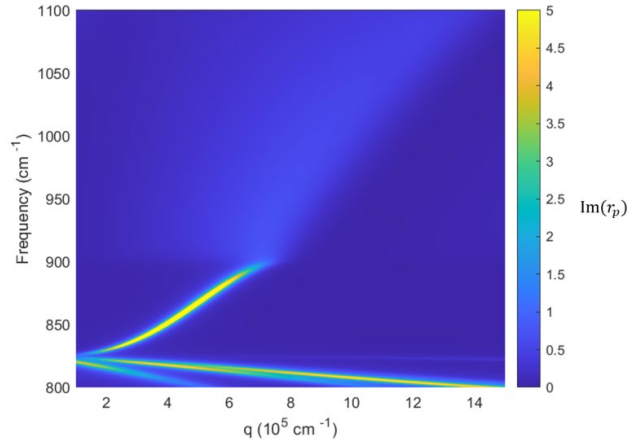

**Figure S4 | Impact of interband transitions in MLG top gate layer on the polaritonic mode.** Plot showing the imaginary part of the reflection coefficient $\text{Im}(r_p)$ for a MLG/hBN (3 nm)/MLG/hBN (30 nm)/SiO$_2$ heterostructure. The Fermi energies of the top and bottom graphene layers are set to 56 meV (equivalent to 450 cm$^{-1}$) and 250 meV respectively.

If the carrier density in the MLG top gate is too low, interband transitions in the MLG top gate can affect nano-infrared imaging experiments. Figure S4 shows the imaginary part of the reflection coefficient $\text{Im}(r_P)$ for a a MLG/hBN (3 nm)/MLG/hBN (30 nm)/SiO$_2$ heterostructure. The Fermi energy of the bottom layer is $E_F^{bot} = 250$ meV which is sufficient to produce a strong plasmonic mode. The Fermi energy of the top layer is $E_F^{top} = 56$ meV which is equivalent in energy to 450cm$^{-1}$. Therefore $\omega_{crit} = 2E_F^{top} = 900\text{cm}^{-1}$ marks the onset of the interband transitions in the top gate and plasmonic mode is damped at higher energies.

Therefore, for experiments performed at $\omega = 905$ cm$^{-1}$, the MLG top gate is not suitable for use in nano-infrared experiments if the Fermi energy and carrier density in the top gate are below the critical values of $E_F^{crit} \sim 56$ meV and $n_{crit} \sim 2.3 \cdot 10^{11}\text{cm}^{-2}$. The rather small value of $n_{crit}$ suggests that this limitation will not be a major hindrance for practical experiments.

Next, we perform simulations to assess the possibility that plasmonic modes in the heterostructure will be scattered by domain walls in twisted bilayer graphene (S10, S13). We use a simplified two-dimensional geometry and compute the electric field in the electrostatic limit using COMSOL software, as shown in Figure S5. We model the tip as a perfectly conducting hyperbola with a minimal radius of curvature of 10 nm. The graphene layers are modeled as 1 nm thick conducting sheets. To approximate the measured signal in nano-infrared experiments, we average the out-of-plane electric field $E_z$ over a 4x4 nm$^2$ area under the tip. We compute the electric fields for several tip positions to obtain a line profile which can be compared with experiment. All calculations are done at a frequency of $\omega = 905$ cm$^{-1}$.

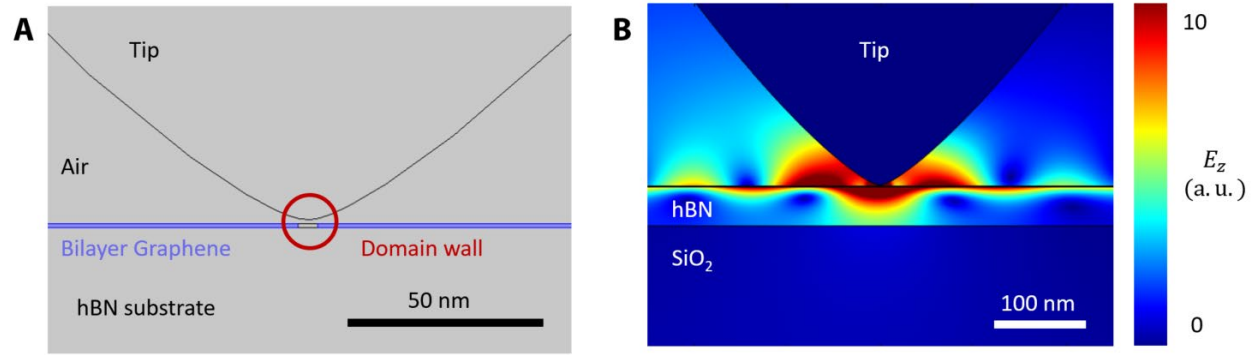

**Figure S5 | Simulation geometry and example of the calculated electric field.** (A) Sketch showing the simplified two dimensional geometry used in our COMSOL simulations for the case where the tip is stationed directly above the domain wall. (B) Typical results of the COMSOL simulation for a heterostructure with no MLG top gate. The oscillatory features in $E_z$ correspond to plasmon polaritons excited by the tip.

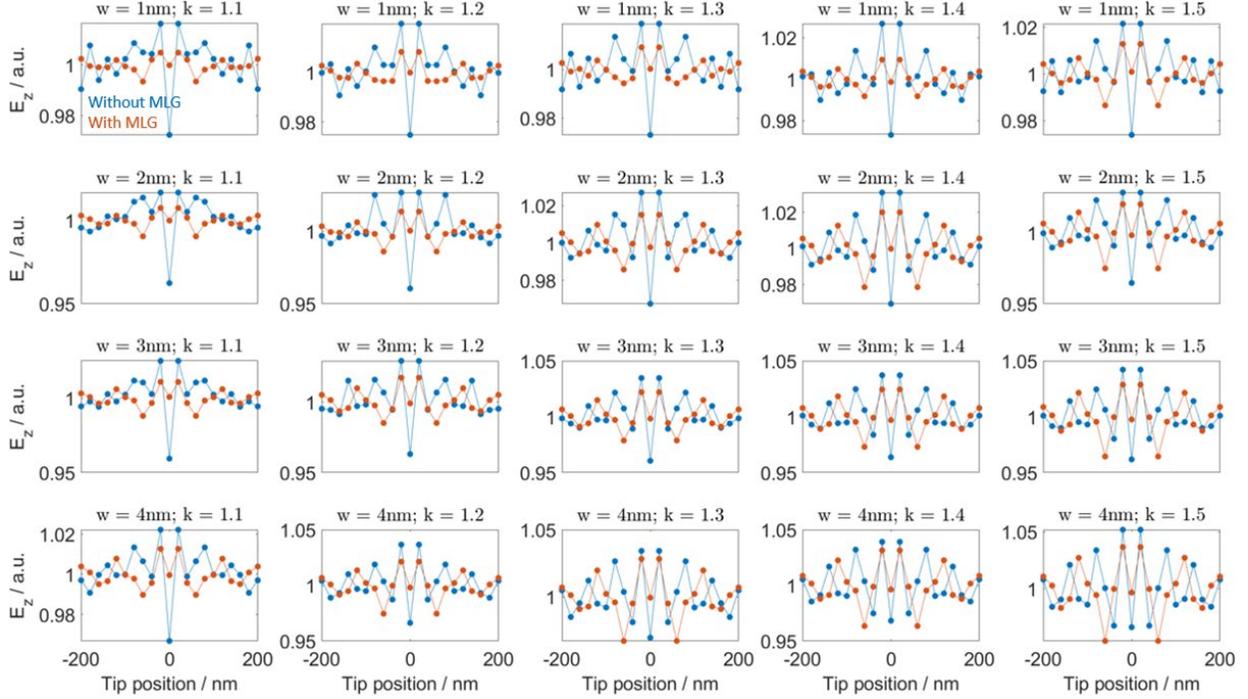

**Figure S6 | Simulations showing plasmon reflection from domain walls in bilayer graphene with and without MLG top gate.** Each panel corresponds to a different value of $w$ and $k$, as indicated. The domain wall is located at the origin of the x-axis.

The conductivity of the graphene sheets is set to the same value of $\sigma_{BLG}$ such that the situation with the MLG top gate corresponds to an equal and opposite carrier density in the MLG and BLG layers. We adjust $\text{Im}(\sigma_{BLG})$ to obtain a plasmon wavelength of ~120 nm with no MLG top gate, a value that's typically observed in experiment (S*10*, S*12*, S*14*) and we set $\text{Im}(\sigma_{BLG})/\text{Re}(\sigma_{BLG}) = 20$ (S*12*). In reality, the conductivity at the domain walls in BLG is anisotropic and displays several additional features (S*10*). However, a rectangular wall can serve as a good approximation (S*10*, S*15*). Therefore, we represent the domain wall as a rectangular region of width $w$ and conductivity $\sigma_{DW} = k\sigma_{BLG}$ where $k$ is a multiplicative factor. We assume that $w \sim 5$ nm and $k \sim 1.5$ are reasonable values based on Figure S5 in Ref (S*10*). Note that larger values of $w$ and $k$ can only lead to stronger plasmonic reflections from the domain wall. We repeat the simulations with and without a MLG top gate and compare the results in Fig S6.

The results in Fig S6 show fringes in $E_z$ as the tip is moved away from the domain wall. The magnitude of these oscillations is ~5% which is comparable to the experimentally observed change in nano-infrared signal without a MLG top gate (S*10*, S*13*). These factors taken together lead us to conclude that our 2D simulations are a good approximation to real experiments. Finally, the magnitude of the plasmonic reflection with the MLG top gate is comparable to the case without a MLG top gate for all values of $w$ and $k$ considered. Therefore, we conclude that domain walls in bilayer graphene could be observed underneath a MLG top gate in future experiments.

**S4.2: Nano-photocurrent simulations: $E_x$ at the graphene layer**

Photocurrent in graphene is generated through the photothermoelectric effect (S16–S18). The absorption of incident light generates hot carriers in graphene. When the hot carriers encounter variations in the Seebeck coefficient, a thermoelectric voltage is generated which drives a current through the sample. Since the electronic conductivity of graphene is negligible in the out-of-plane direction, the in-plane electric field determines the photocurrent response (S19).

Here, we simulate a heterostructure with monolayer graphene with Fermi energy $E_{F,bot} = 250$ meV. We either include or exclude a MLG top gate and compare the electric field profiles. Figure S7(A) shows the absolute value of the in-plane electric field $|E_x|$ and Fig S7(B) shows the line profiles at the probe graphene layer. The addition of the top gate does not significantly affect the in-plane electric field irrespective of its Fermi energy. Therefore, the photocurrent patterns produced by any Seebeck coefficient variations in the bottom graphene layer will be similar to those produced by the same Seebeck coefficient variations in typical nano-photocurrent with no monolayer graphene top gate. These simulations establish that a monolayer graphene top gate can be successfully used for nano-photocurrent experiments under any circumstances.

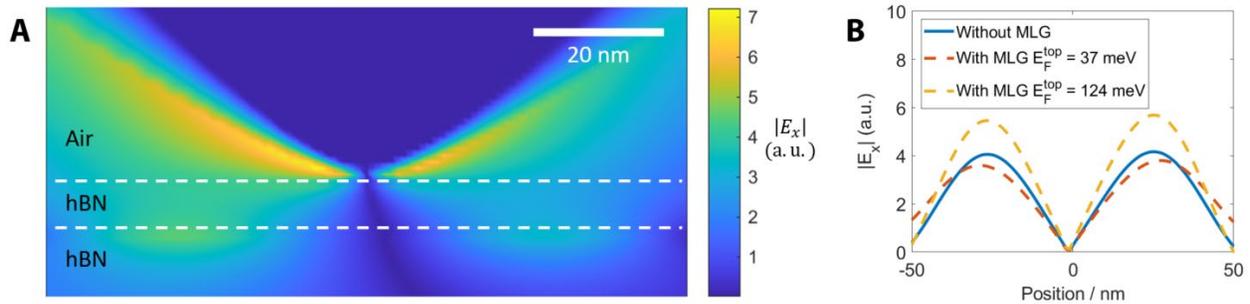

**Figure S7 | In-plane electric field with a MLG top gate.** (A) Two-dimensional plot of the absolute value of the in-plane electric field, $|E_x|$. The two dashed white lines represent the two graphene layers. $E_F^{top} = 37$ meV and $E_F^{bot} = 250$ meV. (B, C) Line profiles of $|E_x|$ for the cases with and without the MLG top gate for two different $E_F^{top}$. $E_F^{bot} = 250$ meV in all cases.

## Section S5: Thickness dependence of the MoS$_2$ top gate

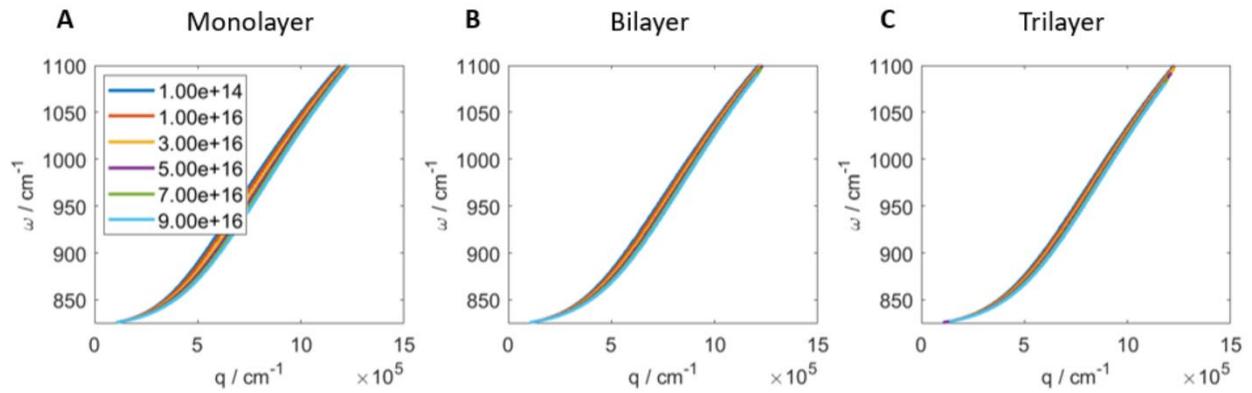

**Figure S8 | Thickness dependence of the MoS$_2$ top gate.** Change in the dispersion of the plasmonic mode of the heterostructure for varying carrier density in the MoS$_2$ top gate. The different panels correspond to different thicknesses of MoS$_2$. The legend corresponds to the carrier densities in the top gate. Panel (B) here is identical to Figure 5(B) of the main text.

In our experiments, we used a bilayer MoS$_2$ as the top gate since it is easier to make electrical contact to multilayer MoS$_2$ than monolayer MoS$_2$ while also minimizing the thickness of the top gate. Here, we explore the effect of varying the MoS$_2$ thickness using the heterostructure considered in Figure 5 of main text. Figure S8 shows the same plot as Figure 5(B) of the main text for three different MoS$_2$ thicknesses. We see that the monolayer MoS$_2$ shows the largest change in the dispersion while the trilayer MoS$_2$ shows the least change. Since the areal carrier density in the top gate is fixed, having more layers means that the charge is spread out over a thicker layer, leading to a smaller volume carrier density and a smaller screening effect.

## Section S6: Gating of bilayer graphene with MoS$_2$

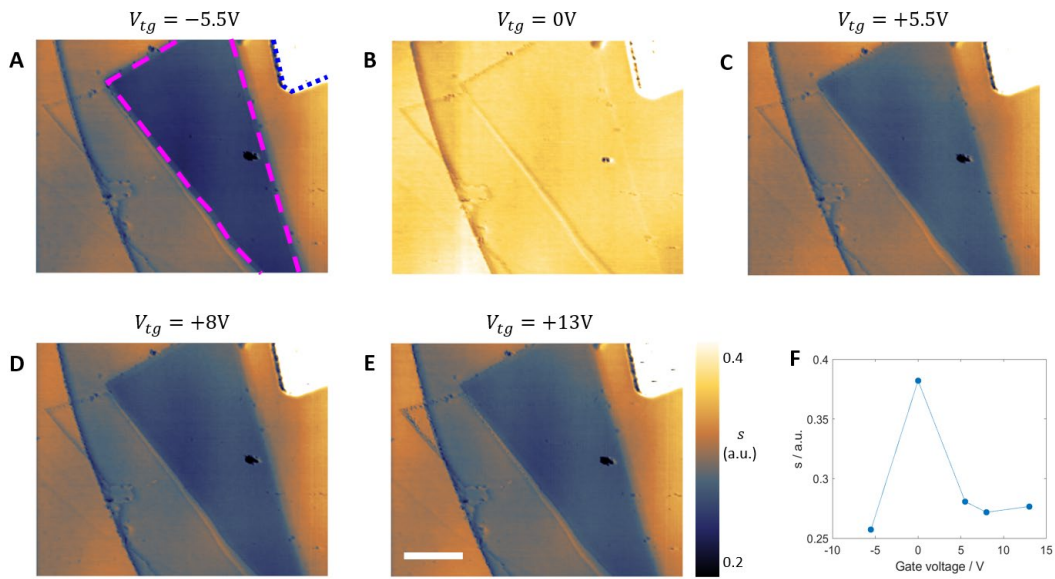

**Figure S9 | Gating bernal bilayer graphene with a MoS$_2$ top-gate (Device 3).** (A) – (E) Nano-infrared amplitude images for $V_{bg} = 0V$ various values of $V_{tg}$. The magenta dashed lines represent the boundaries of the BLG layer while the blue dotted lines represent the boundary of a gold electrode. Scale bar 2μm. (F) Dependence of the nano-infrared amplitude of the bilayer graphene layer on $V_{tg}$. The decrease in the nano-infrared amplitude as $V_{tg}$ is increased is consistent with an increase in the carrier concentration in the BLG layer (S*11*).

In this section, we present nano-infrared data from another bilayer graphene device (Device 3) with a MoS$_2$ top-gate. In this device, there were no domain walls. We observed a change in the nano-infrared signal when varying the top-gate voltage $V_{tg}$.